\documentclass[a4paper]{jpconf}

\usepackage{cite}
\usepackage{graphicx}
\usepackage{epsfig}
\usepackage{dcolumn}
\usepackage{bm}

\begin{document}

\def \d {{\rm d}}
\def \t {{\Theta}}
\def \k {{\kappa}}
\def \l {{\lambda}}
\def \s {{\sigma}}
\def \tr {{\tilde\rho}}
\def \tv {{\tilde v}}
\def \tz {{\tilde z}}
\def \e {{\epsilon}}
\def \P {{p_\lambda}}
\def \Q {{p_\nu}}

\newcommand{\be}{\begin{equation}}
\newcommand{\ee}{\end{equation}}

\newcommand{\beqn}{\begin{eqnarray}}
\newcommand{\eeqn}{\end{eqnarray}}
\newcommand{\AdS}{anti--de~Sitter }
\newcommand{\AAdS}{\mbox{(anti--)}de~Sitter }
\newcommand{\AAN}{\mbox{(anti--)}Nariai }
\newcommand{\AS}{Aichelburg-Sexl }
\newcommand{\ASS}{AS }
\newcommand{\pa}{\partial}
\newcommand{\pp}{{\it pp\,}-}
\newcommand{\ba}{\begin{array}}
\newcommand{\ea}{\end{array}}

\title{Ultrarelativistic boost of spinning and charged black rings\footnote{Proceedings of Fourth Meeting on Constrained Dynamics and Quantum Gravity, Cala Gonone (Sardinia, Italy), September 12--16, 2005. To appear in 
{\em Journal of Physics: Conference Series}.}}

        
\author{Marcello Ortaggio} 
\address{Institute of Theoretical Physics, Faculty of Mathematics and Physics, Charles University in Prague,
         V Hole\v{s}ovi\v{c}k\'{a}ch 2, 180 00 Prague 8, Czech Republic, and INFN, Rome, Italy}
\ead{marcello.ortaggio`AT'comune.re.it}         



\begin{abstract}
We study \AS ultrarelativistic limits in higher dimensions. After reviewing the boost of $D\ge 4$ Reissner-Nordstr{\"{o}}m black holes as a simple illustrative example, we consider the case of $D=5$ black rings, presenting new results for static charged rings.
\end{abstract}
%
%
%





\section{Introduction and review of the boost of $D\ge 4$ Reissner-Nordstr{\"{o}}m black holes}


In 1959 Pirani argued that the geometry associated with a fast moving mass resembles a ``plane'' gravitational wave \cite{piranifast}. Later on, Aichelburg and Sexl (AS) \cite{AicSex71} considered a limiting (``ultrarelativistic'')  boost of the Schwarzschild line element to determine the exact impulsive \pp wave \cite{Penrose68twist} generated by a lightlike particle. In a higher dimensional context, the \ASS limit of static black holes \cite{Tangherlini63} has been known for some time \cite{LouSan90} (also with a dilaton \cite{CaiJiSoh98}). In the case of zero charge, the inclusion of an external magnetic field is straightforward  \cite{Ortaggio05}. The \ASS boost of rotating black holes \cite{MyePer86} has been studied in \cite{Yoshino05}. Here we focus on $D=5$ black rings \cite{EmpRea02prl}. 
In this section we pedagogically review the ultrarelativistic limit of $D\ge 4$ static black holes, as a simple example encompassing all the technical steps. Section~\ref{sec_vacuum} follows \cite{OrtKrtPod05,OrtPodKrt05} in studying the case of vacuum black rings \cite{EmpRea02prl}. In section~\ref{sec_charged} we present new results for static charged rings \cite{IdaUch03,KunLuc05,Yazadjiev05}.

\label{sec_RN}

A static charged black hole is described in any $D\ge 4$ by the line element \cite{Tangherlini63,MyePer86}  
\begin{equation}
 \d s^2=-f^2\d t^2+f^{-2}\d r^2+r^2\d\Omega^2_{D-2} ,
 \label{RN}
\end{equation}
$\d\Omega^2_{D-2}$ being the standard unit ${(D-2)}$-sphere, while $f$ and the electric potential are 
\begin{equation}
 f^2=1-\frac{\mu}{r^{D-3}}+\frac{e^2}{r^{2(D-3)}} , \qquad  A=-\sqrt{\frac{D-2}{2(D-3)}}\frac{e}{r^{D-3}}\,\d t .
 \label{gtt}
\end{equation}

The mass and the charge of the black hole are given by \cite{MyePer86}
\begin{equation}
 M=\frac{\mu(D-2)\Omega_{D-2}}{16\pi} , \qquad Q=e\sqrt{\frac{(D-2)(D-3)}{2}} ,
 \label{MQ}
\end{equation}
and the condition $4e^2\le\mu^2$ must be satisfied in order for the spacetime to be really ``black''. 

For our purposes, it is convenient to decompose the line element~(\ref{RN}) as
\be
 \d s^2=\d s_0^2+\Delta ,
 \label{RNdecomposition}
\ee
in which $\d s_0^2=-\d t^2+\d r^2+r^2\d\Omega^2_{D-2}$ (i.e., eq.~(\ref{RN}) with $\mu=0=e$) is Minkowski spacetime, and 
\be
 \Delta=\left(\frac{\mu}{r^{D-3}}-\frac{e^2}{r^{2(D-3)}}\right)\left(\d t^2+f^{-2}\d r^2\right) .
 \label{RNperturbation}
\ee
Note that for $r\to\infty$ one has $\Delta\to 0$ and $\d s^2\to\d s_0^2$. This enables us to define a Lorentz boost using the symmetries of the flat $\d s_0^2$. We first introduce cartesian coordinates $(z_1,\ldots,z_{D-1})$ via 
\be
 r=\sqrt{z_1^2+\rho^2} , \qquad \rho=\sqrt{z_iz^i} \quad (i=2,\ldots,D-1) , 
 \label{RNradius}
\ee
and then suitable double null coordinates $(u',v')$ by
\begin{equation}
 t=\frac{-u'+v'}{\sqrt{2}} , \qquad z_1=\frac{u'+v'}{\sqrt{2}} .
 \label{nullcoords}
\end{equation}
Now, a boost along the (generic) direction $z_1$ takes the simple form
\begin{equation}
 u'=\epsilon^{-1}u  , \qquad v'=\epsilon v  .
 \label{lorentzboost}
\end{equation}
An ``ultrarelativistic''
boost to the speed of light amounts in taking $\e\to 0$ in eq.~(\ref{lorentzboost}). We will also rescale the mass and the electric charge 
as $M=\gamma^{-1}p_M$, $Q^2=\gamma^{-1}p_Q^2$ \cite{AicSex71,LouSan90}, i.e. (for $\e\to 0$) 
\be 
 M\approx2\e p_M , \qquad Q^2\approx2\e p_Q^2 \qquad (p_M,p_Q>0).
 \label{RNrescalings}
\ee

We can now make the substitutions~(\ref{RNradius})--(\ref{RNrescalings}) in the black hole metric~(\ref{RN}) (i.e., eqs.~(\ref{RNdecomposition}) and (\ref{RNperturbation})). The flat background becomes $\d s_0^2=2\d u\d v+\d z_i\d z^i$, and $\Delta$ depends parametrically on $\e$
\beqn
 \Delta_\e= & & \left(\frac{32\pi}{(D-2)\Omega_{D-2}}\frac{p_M}{\tilde r^{D-3}}-\frac{4}{(D-2)(D-3)}\frac{p_Q^2}{\tilde r^{2(D-3)}}\right) \nonumber \\ & & {}\times   
  \e\left[\frac{1}{2}(-\e^{-1}\d u+\e\d v)^2+f^{-2}(\tilde r)\left(\frac{z_\e}{\tilde r}\frac{1}{\sqrt{2}}(-\e^{-1}\d u+\e\d v)+\frac{z_i}{\tilde r}\d z^i\right)^2\right] ,
 \label{RNperturbation2}
\eeqn
\be
 \mbox{where } \qquad \tilde r=\tilde r(z_\e,z_i)=\sqrt{z_\e^2+\rho^2} , \qquad z_\e=\frac{1}{\sqrt{2}}(\e^{-1}u+\e v) .
 \label{shortcut}
\ee
We have introduced the combination $z_\e$ because it fully determines how ${\Delta_\e}$ depends on ${u}$. In the limit ${\d s^2=\d s_0^2+\lim_{\e\to 0}\Delta_\e}$, the expansion of ${\Delta_\e}$ in ${\e}$ displays a peculiar behaviour at ${u=0}$
\be
 \Delta_\e=\frac{1}{\e}h(z_\e)\d u^2+\dots ,
 \label{dominant}
\ee
\be
 h(z_\e)=\frac{32\pi}{(D-2)\Omega_{D-2}}\frac{p_M}{2}\left(\frac{1}{\tilde r^{D-3}}+\frac{z_\e^2}{\tilde r^{D-1}}\right)-\frac{4}{(D-2)(D-3)}\frac{p_Q^2}{2}\left(\frac{1}{\tilde r^{2(D-3)}}+\frac{z_\e^2}{\tilde r^{2(D-2)}}\right) ,
 \label{hRN}
\ee
and the dots denote terms which become negligible in the limit. We have emphasized here the essential dependence of the function $h$ on $z_\e$ (and thus on $\e$), but of course $h$ depends also on $\rho$. In taking the limit $\e\to 0$ of eq.~(\ref{dominant}), we apply the distributional identity
\be
 \lim_{\e\to 0} \frac{1}{\e}g\left(z_\e \right)
  =\sqrt{2}\,\delta(u)\int_{-\infty}^{+\infty}g(z)\d z .
 \label{identity}
\ee
The final metric is thus 
\be
 \d s^2=2\d u\d v+\d z_i\d z^i+H\delta(u)\d u^2 ,
 \label{ppgeneral}
\ee 
\be
 \mbox{with } \qquad H=\sqrt{2}\int_{-\infty}^{+\infty}h(z)\d z ,
 \label{integ}
\ee
and eq.~(\ref{hRN}). For any $D>4$ one can readily employ the integral $2\int_0^\infty x^{2a-1}/(1+x^2)^{a+b}\d x=\Gamma(a)\Gamma(b)/\Gamma(a+b)$ (with appropriate values $a,b>0$) to evaluate~(\ref{integ}), and one obtains explicitly 
\beqn
 H= & & C_M\frac{p_M}{\rho^{D-4}}
      -C_Q\frac{p_Q^2}{\rho^{2(D-3)-1}} \qquad (D>4) , \nonumber \label{HD} \\
 \mbox{ where }  & & \quad  C_M=\frac{16\pi\sqrt{2}}{(D-4)\Omega_{D-3}} , \qquad   
 C_Q=\frac{(2D-9)!!}{(D-3)!}\frac{2D-5}{(D-2)(D-3)}\frac{\pi\sqrt{2}}{2^{D-4}} .
\eeqn

For $D=4$ the integral of (\ref{hRN}) is divergent, due (only) to the terms proportional to $p_M$. One can perform the (by now standard) infinite ``gauge'' subtraction of \cite{AicSex71}\footnote{Ref.~\cite{BalNac95} pointed out that such a procedure contains some ambiguity, and presented a  different approach based on boosting the energy-momentum tensor. In the case, e.g., of the boosted Schwarzschild metric, one can alternatively remove the ambiguity by using a simple symmetry argument, since the final \pp wave must be an axially symmetric vacuum solution (which uniquely determines the result of \cite{AicSex71}). No ambiguity arises for $D>4$.} to eventually obtain
\be
 H=-8\sqrt{2}p_M\ln\rho-\frac{3\pi\sqrt{2}}{2}\frac{p_Q^2}{\rho} \qquad (D=4) .
 \label{H4}
\ee 

A $D$-dimensional charged black hole boosted to the speed of light is
thus described by the impulsive \pp wave~(\ref{ppgeneral}) with eqs.~(\ref{HD}) or (\ref{H4}). For $p_Q=0$, this is just the (higher dimensional) \ASS vacuum solution. The spacetime~(\ref{ppgeneral}) is flat everywhere except on the  wave front $u=0$. There is a singularity at $\rho=0$, as a remnant of the curvature singularity $r=0$ of the original metric~(\ref{RN}). Because of eq.~(\ref{RNrescalings}) for $Q$, the Maxwell field ${F=\d A}$ (cf.~eq.~(\ref{gtt})) tends to zero when $\e\to 0$, but its energy-momentum tensor is non-vanishing and proportional to $p_Q^2\delta(u)$ \cite{LouSan90} (such a physically ``peculiar configuration'' is, however, mathematically sound \cite{Steinbauer97}). On the other hand, if one replaced the rescaling~(\ref{RNrescalings}) by $Q=\mbox{const.}$, one would end up with a non-zero $F_{\mu\nu}\sim Q\delta(u)$, but with an ill-defined $T_{\mu\nu}\sim Q^2\delta^2(u)$. Note also that the rescalings~(\ref{RNrescalings}) imply a violation of the charge bound $4e^2\le\mu^2$ for a small $\e$. The \pp wave presented above thus corresponds to the boost of a charged naked singularity rather than a black hole. In order to preserve the event horizon(s) until the very final limit, one should rescale $Q$ (i.e., $e$) at least as fast as $Q^2\sim\e^2$, but one would then simply recover the \ASS vacuum geometry \cite{LouSan90,CaiJiSoh98}. 
See, e.g., the reviews \cite{Podolsky02} and references therein for other \ASS limits and for general properties of impulsive waves.

\section{Boost of the vacuum black ring}

\label{sec_vacuum}

We now consider the \ASS boost of the black ring \cite{EmpRea02prl}. We start from the metric form of \cite{Emparan04}
\beqn
 \d s^2= & & -\frac{F(y)}{F(x)}\left(\d t+C(\nu,\lambda)L\frac{1+y}{F(y)}\d\psi\right)^2 \nonumber \label{ring} \\
 & & {}+\frac{L^2}{(x-y)^2}F(x)\left[-\frac{G(y)}{F(y)}\d\psi^2-\frac{\d y^2}{G(y)}+\frac{\d x^2}{G(x)}+\frac{G(x)}{
      F(x)}\d\phi^2\right] ,
\eeqn
\be
 \mbox{where } \quad F(\zeta)=\frac{1+\lambda\zeta}{1-\lambda} , \quad G(\zeta)=(1-\zeta^2)\frac{1+\nu\zeta}{1-\nu} , \quad    
 C(\nu,\lambda)=\sqrt{\frac{\lambda(\lambda-\nu)(1+\lambda)}{(1-\nu)(1-\lambda)^3}} .
 \label{FG}
\ee
The black ring~(\ref{ring}) is asymptotically flat near spatial infinity $x,y\to -1$ (see \cite{EmpRea02prl,Emparan04} for more details). Mass, angular momentum and angular velocity (at the horizon) are 
\be
 M=\frac{3\pi L^2}{4}\frac{\lambda}{1-\lambda} , \qquad   
 J=\frac{\pi L^3}{2}\sqrt{\frac{\lambda(\lambda-\nu)(1+\lambda)}{(1-\nu)(1-\lambda)^3}} , \qquad   
 \Omega=\frac{1}{L}\sqrt{\frac{(\lambda-\nu)(1-\lambda)}{\lambda(1+\lambda)(1-\nu)}} .
 \label{mass}
\ee
With the choice $\psi, \phi\in[0,2\pi]$ there are no conical singularities at $y=-1$ and $x=-1$. The black ring is ``in equilibrium'' if conical singularities are absent also at $x=+1$, which requires 
\be
 \lambda=\frac{2\nu}{1+\nu^2} .
 \label{equilibrium}
\ee

To perform the \ASS limit of the black ring, we decompose eq.~(\ref{ring}) as in eq.~(\ref{RNdecomposition}), where the flat $\d s_0^2$ is now given by eq.~(\ref{ring}) with $\lambda=0=\nu$, and $\Delta$ is a cumbersome expression given in \cite{OrtPodKrt05}. In order to express the boosts of $\d s_0^2$, we first define new coordinates $(\xi,\eta)$ via 
\be
 y=-\frac{\xi^2+\eta^2+L^2}{\sqrt{(\eta^2+\xi^2-L^2)^2+4L^2\eta^2}} \, , \qquad
 x=-\frac{\xi^2+\eta^2-L^2}{\sqrt{(\eta^2+\xi^2-L^2)^2+4L^2\eta^2}} \, .
 \label{cylindrical}
\ee 
Then, spatial cartesian coordinates $(x_1,x_2,y_1,y_2)$ are given by
\be
 x_1=\eta\cos\phi , \quad x_2=\eta\sin\phi , \qquad \qquad y_1=\xi\cos\psi , \quad y_2=\xi\sin\psi ,
 \label{cartesian}
\ee 
so that $\d s_0^2=-\d t^2+\d x_1^2+\d x_2^2+\d y_1^2+\d y_2^2$. This enables us to study a
boost along a general direction $z_1$, which can be specified by a single parameter~$\alpha$ if we introduce rotated axes $(z_1,z_2)$ 
\be
 x_1=z_1\cos\alpha-z_2\sin\alpha , \qquad y_1=z_1\sin\alpha+z_2\cos\alpha .
 \label{rotated}
\ee 
Using null coordinates as in eq.~(\ref{nullcoords}), we can finally perform the boost~(\ref{lorentzboost}).
In the ultrarelativistic limit $\e\to 0$, we again rescale the mass as $M=\gamma^{-1}p_M\approx2\e p_M$. In addition, we wish to keep the angular velocity $\Omega$ finite. From eq.~(\ref{mass}), these requirements suggest the rescalings
\be
 \lambda=\epsilon \P , \qquad  \nu=\epsilon \Q ,
 \label{ASlambda}
\ee where $p_\lambda=8p_M/(3\pi L^2)$, and $p_\nu$ is another positive constant
such that $\P\ge\Q$. In terms of these parameters, for $\e\to 0$ the
{\em equilibrium condition}~(\ref{equilibrium}) becomes $\P=2\Q$.

All the ingredients necessary to evaluate how the metric~(\ref{ring}) transforms under the boost~(\ref{lorentzboost}) have been given above, and we skip intermediate lengthy calculations (see \cite{OrtKrtPod05,OrtPodKrt05} for more details). As in section~\ref{sec_RN}, after the splitting~(\ref{RNdecomposition}) one obtains for $\Delta$ an expression of the form~(\ref{dominant}) when $\e\to 0$. With eq.~(\ref{identity}), one finds that the final metric is a $D=5$ impulsive \pp wave $\d s^2=2\d u\d v+\d x_2^2+\d y_2^2+\d z_2^2+H(x_2,y_2,z_2)\delta(u)\d u^2$, with eq.~(\ref{integ}). The integral~(\ref{integ}) can be expressed explicitly in terms of elliptic integrals at least for boosts along the privileged axes $x_1$ ($\alpha=0$) and $y_1$ ($\alpha=\pi/2$), respectively ``orthogonal'' and ``parallel'' to the 2-plane $(y_1,y_2)$.

For an {\em orthogonal} boost ($\alpha=0$) the final \pp wave, propagating along $x_1$, reads
\be
 \d s^2=2\d u\d v+\d x_2^2+\d y_1^2+\d y_2^2+H_{_{\!\bot}}\!(x_2,y_1,y_2)\delta(u)\d u^2 ,  
 \label{pporth}
\ee
\beqn
 H_{_{\!\bot}}= & & \sqrt{2}\,\frac{3\P L^2+(2\Q-\P)\xi^2}{\sqrt{(\xi+L)^2+x_2^2}}\,K(k)+\sqrt{2}(2\Q-\P) \nonumber \label{Horth} \\ 
 & & {}\times \left[-\sqrt{(\xi+L)^2+x_2^2}\,E(k)+
      \frac{\xi-L}{\xi+L}\frac{x_2^2}{\sqrt{(\xi+L)^2+x_2^2}}\,\Pi(\rho,k)+\pi|x_2|\Theta(L-\xi)\right] , 
\eeqn
in which
\be
 k=\sqrt{\frac{4\xi L}{(\xi+L)^2+x_2^2}} , \qquad \rho=\frac{4\xi L}{(\xi+L)^2} , \qquad \xi=\sqrt{y_1^2+y_2^2} ,
 \label{krho_orth}
\ee
and $\Theta(L-\xi)$ denotes the step function (cf.~the appendix of \cite{OrtKrtPod05} for definitions of the elliptic integrals $K$, $E$ and $\Pi$). The function~(\ref{Horth}) simplifies considerably for the more interesting case of black rings in equilibrium, i.e. $\P=2\Q$. This corresponds to a spacetime which is vacuum everywhere except on the singular circle at $u=0=x_2$, $\xi=L$, a remnant of the curvature singularity ($y=-\infty$) of the original black ring (\ref{ring}). For $\P\neq 2\Q$, the discontinuous term proportional to $\Theta(L-\xi)$ is responsible for a disk membrane supporting the ring.

In the case of a {\em parallel} boost ($\alpha=\pi/2$) the final \pp wave, now propagating along $y_1$, is
\be
 \d s^2=2\d u\d v+\d x_1^2+\d x_2^2+\d y_2^2+H_{_{\!||}}\!(x_1,x_2,y_2)\delta(u)\d u^2 , 
 \label{ppparall}
\ee
\beqn
 & & H_{_{\!||}}= \left[2(2\P-\Q)L^2+(2\Q-\P)a^2\left(1+\frac{L^2+\eta^2}{a^2-y_2^2}\right)\right. \nonumber \\ & & \hspace{1.5cm} \left.+2\sqrt{\P(\P-\Q)}Ly_2\left(1-\frac{L^2+\eta^2}{a^2-y_2^2}\right)\right]
\frac{\sqrt{2}}{a}K(k')-2\sqrt{2}(2\Q-\P)aE(k') \nonumber \label{Hparall} \\ 
 & & \quad                
 {}+\frac{\sqrt{2}}{2}\left[(2\Q-\P)y_2-2\sqrt{\P(\P-\Q)}L\right]\left[-\frac{\eta^2+L^2}{ay_2}\frac{a^2+y_2^2}{a^2-y_2^2}\,
    \Pi(\rho',k')+\pi\,\mbox{sgn}(y_2)\right] , 
\eeqn
where
\beqn
 & & k'=\frac{\left(a^2-\eta^2-y_2^2+L^2\right)^{1/2}}{\sqrt{2}a} , \qquad     
      \rho'=-\frac{(a^2-y_2^2)^2}{4a^2y_2^2} , \nonumber  \label{krho_parall} \\
 & & a=\left[(\eta^2+y_2^2-L^2)^2+4\eta^2L^2\right]^{1/4} , \qquad \eta=\sqrt{x_1^2+x_2^2} .
\eeqn
The profile~(\ref{Hparall}) is simpler for $\P=2\Q$, when it describes a vacuum spacetime singular on the rod at $u=0=\eta$ and $|y_2|\le L$. In \cite{OrtPodKrt05} we also pointed out distinct features with respect to boosted black holes \cite{Yoshino05}, and we analyzed the \ASS limit of the supersymmetric black ring of \cite{Elvangetal04}. 

\section{Boost of the charged black ring}

\label{sec_charged}

We present here new results describing the \ASS boost of static charged black rings. These were found in \cite{IdaUch03} (up to a misprint in $F_{\mu\nu}$) in the Einstein-Maxwell theory, generalized to dilaton gravity in \cite{KunLuc05}, and rederived more systematically in \cite{Yazadjiev05}. The solution of \cite{IdaUch03} coincides with the static limit of certain supergravity black rings \cite{ElvEmp03}. It can also be obtained by Harrison-transforming (along $\pa_t$) a static vacuum black ring (eq.~(\ref{ring}) with $\nu=\lambda$), which  gives
\be
 \d s^2=-\Lambda^{-2}\frac{F(y)}{F(x)}\d t^2+\frac{\Lambda L^2}{(x-y)^2}F(x)\bigg[(y^2-1)\d\psi^2+\frac{1}{F(y)}\frac{\d y^2}{y^2-1}+\frac{1}{F(x)}\frac{\d x^2}{1-x^2}+(1-x^2)\d\phi^2\bigg] ,
 \label{ring_el}
\ee
with $F(\zeta)$ as in eq.~(\ref{FG}). The function $\Lambda$ and the electromagnetic potential are given by 
\be
 \Lambda=\frac{1-e^2\frac{F(y)}{F(x)}}{1-e^2} , \qquad A=-\frac{\sqrt{3}\,e}{2\Lambda}\frac{F(y)}{F(x)}\,\d t ,
\ee
and the parameter $e$ satisfies $e^2<1$. Mass and electric charge of the black ring are
\be
 M=\frac{3\pi L^2}{4}\frac{\lambda}{1-\lambda}\frac{1+e^2}{1-e^2} , \qquad        
 Q=2L^2\frac{\lambda}{1-\lambda}\frac{\sqrt{3}\,e}{1-e^2} .
 \label{mass_charge}
\ee
In order to perform the ultrarelativistic boost, one can follow the method outlined in sections~\ref{sec_RN} and \ref{sec_vacuum}. For $\e\to0$ we rescale the mass as $M\approx2\e p_M$, that is $\lambda=\e\P$ as in eq.~(\ref{ASlambda}) (hence $\P=p_M\frac{8}{3\pi L^2}(1-e^2)/(1+e^2)$), and we keep $e$ constant. This will automatically enforce $Q\sim\e$, which is different from the rescaling~(\ref{RNrescalings}) for black holes (note that $M/Q=\frac{\sqrt{3}\pi}{8}(1+e^2)/e$ here). Referring to the key coordinate transformations~(\ref{cylindrical})--(\ref{rotated}) of section~\ref{sec_vacuum}, we directly jump to the final result. 
For an {\em orthogonal} boost we obtain an impulsive \pp wave of the form~(\ref{pporth}), with
\beqn
 H^c_{_{\!\bot}}= & & \sqrt{2}\P\Bigg[\left(3L^2\frac{1+e^2}{1-e^2}+\xi^2+x_2^2\frac{\xi+L}{\xi-L}\right)\frac{K(k)}{\sqrt{(\xi+L)^2+x_2^2}} \nonumber  \label{Horth_el} \\ 
 & & {}-\sqrt{(\xi+L)^2+x_2^2}\,E(k) -\frac{\xi+L}{\xi-L}\frac{(\xi-L)^2+x_2^2}{\sqrt{(\xi+L)^2+x_2^2}}\,\Pi(\rho_0,k)+\frac{\pi}{2}|x_2| \Bigg] , 
\eeqn
$k$ and $\xi$ as in eq.~(\ref{krho_orth}) and $\rho_0=-(\xi-L)^2/x_2^2$.
For a {\em parallel} boost, one finds a metric~(\ref{ppparall}) with
\beqn
 H^c_{_{\!||}}= & & \sqrt{2}\P\left[\left(2L^2\frac{1+2e^2}{1-e^2}+ a^2+a^2\frac{L^2+\eta^2}{a^2-y_2^2}\right)
\frac{1}{a}K(k')-2aE(k')\right. \nonumber \label{Hparall_el} \\ 
 & & {}\left.-\frac{\eta^2+L^2}{2a}\frac{a^2+y_2^2}{a^2-y_2^2}\,
    \Pi(\rho',k')+\frac{\pi}{2}|y_2|\right] , 
\eeqn
where $k'$, $\rho'$, $a$ and $\eta$ are as in eq.~(\ref{krho_parall}).
The above profiles reduce to those for boosted static {\em vacuum} black rings \cite{OrtKrtPod05} when $e=0$. The function $H^c_{_{\!\bot}}$ is singular on the circle $u=0=x_2$, $\xi=L$, which is the boundary of a disk-shaped membrane (a remnant of the conical singularity of the spacetime~(\ref{ring_el})). The function $H^c_{_{\!||}}$ is singular on the rod $u=0=\eta$, $|y_2|\le L$, where the membrane has also contracted now. The rescaling $Q\sim\e$ (discussed above) implies that $H^c_{_{\!\bot}}$ and $H^c_{_{\!||}}$ represents {\em vacuum solutions}, i.e. both $F_{\mu\nu}$ and $T_{\mu\nu}$ tend to zero for $\e\to 0$. Note, however, that in eqs.~(\ref{Horth_el}) and (\ref{Hparall_el}) there is still a signature of the original electric charge (via $e$). This enables us to perform the ``BPS limit'' \cite{IdaUch03,KunLuc05} $e^2\to 1^-$ (simultaneously, replace $\P$ with $p_M$) of $H^c_{_{\!\bot}}$ and $H^c_{_{\!||}}$, which leads to simple expressions with no membrane disk. Exactly the same expressions can be recovered from the boost of the supersymmetric black ring \cite{OrtPodKrt05}.

\ack

The author is supported by a post-doctoral fellowship from INFN (bando n.10068/03).

\addvspace{.3cm}

%

\begin{thebibliography}{10}

\bibitem{piranifast}
Pirani F~A~E 1959 {\em Proc. R. Soc. {\rm A}\/} {\bf 252} 96

\bibitem{AicSex71}
Aichelburg P~C and Sexl R~U 1971 {\em Gen. Rel. Grav.\/} {\bf 2} 303

\bibitem{Penrose68twist}
Penrose R 1968 {\em Int. J. Theor. Phys.\/} {\bf 1} 61;
1972 {\em General Relativity: Papers in Honour of J. L. Synge\/}, ed
  L~O'Raifeartaigh (Oxford: Clarendon Press) pp. 101--115

\bibitem{Tangherlini63}
Tangherlini F~R 1963 {\em Nuovo Cimento\/} {\bf 27} 636

\bibitem{LouSan90}
Loust\'o C~O and S\'anchez N 1990 {\em Int. J. Mod. Phys. {\rm A}\/} {\bf 5}
  915

\bibitem{CaiJiSoh98}
Cai R~G, Ji J~Y and Soh K~S 1998 {\em Nucl. Phys. {\rm B}\/} {\bf 528} 265

\bibitem{Ortaggio05}
Ortaggio M 2005 {\em JHEP\/} {\bf 05} 048 [gr-qc/0410048]

\bibitem{MyePer86}
Myers R~C and Perry M~J 1986 {\em Ann. Phys. (N.Y.)\/} {\bf 172} 304

\bibitem{Yoshino05}
Yoshino H 2005 {\em Phys. Rev. {\rm D}\/} {\bf 71} 044032

\bibitem{EmpRea02prl}
Emparan R and Reall H~S 2002 {\em Phys. Rev. Lett.\/} {\bf 88} 101101

\bibitem{OrtKrtPod05}
Ortaggio M, Krtou{\v s} P and Podolsk\'y J 2005 {\em Phys. Rev. {\rm D}\/} {\bf
  71} 124031

\bibitem{OrtPodKrt05}
Ortaggio M, Podolsk\'y J and Krtou{\v s} P 2005 {\em JHEP\/} {\bf 12} 001 [gr-qc/0506064]

\bibitem{IdaUch03}
Ida D and Uchida Y 2003 {\em Phys. Rev. {\rm D}\/} {\bf 68} 104014

\bibitem{KunLuc05}
Kunduri H~K and Lucietti J 2005 {\em Phys. Lett. {\rm B}\/} {\bf 609} 143

\bibitem{Yazadjiev05}
Yazadjiev S~S 2005  [{h}ep-th/0507097]

\bibitem{BalNac95}
Balasin H and Nachbagauer H 1995 {\em Class. Quantum Grav.\/} {\bf 12} 707

\bibitem{Steinbauer97}
Steinbauer R 1997 {\em J. Math. Phys.\/} {\bf 38} 1614

\bibitem{Podolsky02}
Podolsk\'y J 2002 {\em Gravitation: Following the Prague Inspiration\/}, eds
  O~Semer\'ak, J~Podolsk\'y and M~\v{Z}ofka (Singapore: World Scientific) pp.
  205--246 [gr--qc/0201029]; 
Ortaggio M 2002 {\em On Some Properties of Impulsive Waves in {G}eneral
  {R}elativity\/} {PhD} thesis Universit\`a degli {S}tudi di {T}rento
  [available at {\tt http://utf.mff.cuni.cz/ortaggio/}]

\bibitem{Emparan04}
Emparan R 2004 {\em JHEP\/} {\bf 03} 064 [hep-th/0402149]

\bibitem{Elvangetal04}
Elvang H, Emparan R, Mateos D and Reall H~S 2004 {\em Phys. Rev. Lett.\/} {\bf
  93} 211302

\bibitem{ElvEmp03}
Elvang H and Emparan R 2003 {\em JHEP\/} {\bf 11} 035 [hep-th/0310008]; 
Elvang H, Emparan R and Figueras P 2005 {\em JHEP\/} {\bf 02} 031 [hep-th/0412130]

\end{thebibliography}

\end{document}